\documentclass[showpacs,twocolumn,floats,superscriptaddress]{revtex4}
\usepackage{graphicx}
\usepackage{amsmath}
 
\newcommand{\bq}{\begin{equation}} 
\newcommand{\ee}{\end{equation}}

\begin{document}
\title{
Noiseless scattering states in a chaotic cavity
}
\author{P. G. Silvestrov}
\affiliation{Instituut-Lorentz, Universiteit Leiden, P.O. Box 9506, 2300 RA
Leiden, The Netherlands}
\affiliation{Budker Institute of Nuclear Physics, 630090 Novosibirsk, Russia}
\author{M. C. Goorden}
\affiliation{Instituut-Lorentz, Universiteit Leiden, P.O. Box 9506, 2300 RA
Leiden, The Netherlands}
\author{C. W. J. Beenakker}
\affiliation{Instituut-Lorentz, Universiteit Leiden, P.O. Box 9506, 2300 RA
Leiden, The Netherlands}
\date{11 February 2003}
\begin{abstract}
Shot noise in a chaotic cavity (Lyapunov exponent $\lambda$, level
spacing $\delta$, linear dimension $L$), coupled by two $N$-mode point
contacts to electron reservoirs, is studied as a measure of the crossover
from stochastic quantum transport to deterministic classical transport.
The transition proceeds through the formation of {\em fully\/} 
transmitted or reflected scattering states, which we construct explicitly.
The fully transmitted states contribute
to the mean current $\bar{I}$, but not to the shot-noise power $S$.
We find that these noiseless transmission channels do not exist for 
$N\alt\sqrt{k_{F}L}$, where we expect the random-matrix result
$S/2e\bar{I}=1/4$. For $N\agt\sqrt{k_{F}L}$ we predict a 
suppression of the noise $\propto (k_{F}L/N^{2})^{N\delta/\pi\hbar\lambda}$.
This nonlinear contact dependence of the noise could help to distinguish
ballistic chaotic scattering from random impurity scattering in quantum
transport.

\end{abstract}
\maketitle

Shot noise can distinguish deterministic scattering, characteristic for
particles, from stochastic scattering, characteristic for waves. Particle
dynamics is deterministic: A given initial position and momentum fixes
the entire trajectory. In particular, it fixes whether the particle
will be transmitted or reflected, so the scattering is noiseless. Wave
dynamics is stochastic: The quantum uncertainty in position and momentum
introduces a probabilistic element into the dynamics, so it is noisy.

The suppression of shot noise in a conductor with deterministic
scattering was predicted many years ago from this qualitative argument
\cite{Bee91}. A better understanding, and a quantitative description,
of how shot noise measures the transition from particle to wave dynamics
in a chaotic quantum dot was put forward by Agam, Aleiner, and Larkin
\cite{Aga00}, and developed further in Ref.~\cite{Hen02}. 
The key concept is the Ehrenfest time $\tau_{E}$, which
is the characteristic time scale of quantum chaos \cite{Zas81}. The
noise power $S\propto\exp(-\tau_{E}/\tau_{ D})$ was predicted to
vanish exponentially with the ratio of $\tau_{E}$ and the mean dwell time
$\tau_{D}=\pi\hbar/N\delta$ in the quantum dot (with $\delta$
the level spacing and $N$ the number of modes in each of the two point
contacts through which the current is passed). A recent measurement of
the $N$-dependence of $S$ is consistent with this prediction for
$\tau_{E}<\tau_{ D}$, although
an alternative explanation in terms of short-range impurity scattering
describes the data equally well \cite{Obe02}.

\begin{figure}
\includegraphics[width=6.5cm]{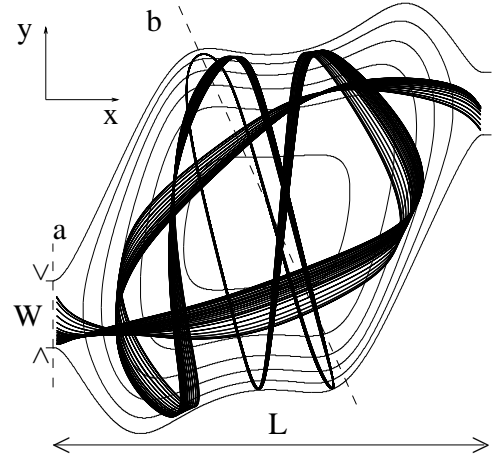}
\caption{ Selected equipotentials of the electron billiard. The outer
equipotential is at $E_{F}$, the other equipotentials are at increments
of $0.19\, E_{F}$. Dashed lines $a$ and $b$ show the sections described 
in the text. Also shown is a
flux tube of transmitted trajectories, all originating from a
single closed contour in a transmission band, representing the spatial
extension of a fully transmitted scattering state. The flux tube is wide
at the two openings and squeezed inside the billiard.}
\label{figure1}
\vspace{-.2cm}
\end{figure}

The theory of Ref.\ \cite{Aga00} introduces the stochastic element by
means of long-range impurity scattering and adjusts the scattering rate
so as to mimic the effect of a finite Ehrenfest time. Here we take the 
alternative approach of explicitly constructing noiseless channels
in a chaotic quantum dot. These are scattering states which are either
fully transmitted or fully reflected in the semiclassical limit. 
They are not described by random matrix theory~\cite{Jal94}. 
By determining what fraction of the available
channels is noiseless, we can deduce a precise upper bound for the 
shot-noise power. A random matrix conjecture for the remaining noisy
channels gives an explicit form of $S(N)$.
We find that the onset of the classical suppression of the noise
is described not only by the Ehrenfest time, but by the difference
of $\tau_{E}$ and the ergodic time $\tau_0$, which we introduce and
calculate in the paper. The resulting nonlinear dependence of 
$\ln S$ on $N$ may help to distinguish between
the competing explanations of the experimental data~\cite{Obe02}.

\begin{figure}
\includegraphics[width=7.cm]{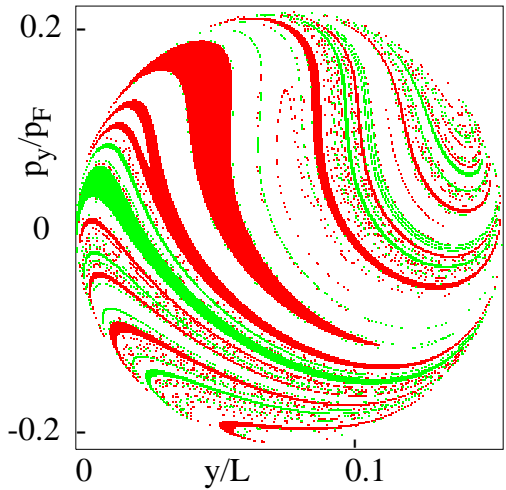}
\caption{
Section of phase space at $p_{x}=\sqrt{p_{F}^{2}-p_{y}^{2}}$ and $x=0$, 
corresponding to line $a$ in Fig.~1. Each dot in this surface of section is
the starting point of a classical trajectory that is transmitted through
the lead at $x=L$ (black/red), or reflected back through $x=0$ (gray/green). 
The points lie in narrow bands. Only the trajectories with dwell time
$t<12mL/p_F$ are shown.
%precisely: $t<12.2mL/p_F$ Reminder --- $p_W=.2p_F$, $W=.9$,
%%$W_L=.93$, $W_R=.87$, $L=6.$, $m=1$.
}
\label{figure2}
\vspace{-.2cm}
\end{figure}

We illustrate the construction of noiseless scattering states for the
two-dimensional billiard with smooth confining potential $U(x,y)$
shown in Fig.~1. The outer equipotential defines the
area in the $x-y$ plane which is classically accessible at the Fermi
energy $E_{F}=p_{F}^{2}/2m$ (with $p_{F}=\hbar k_{F}$ the Fermi
momentum). The motion in the closed billiard is chaotic with Lyapunov
exponent $\lambda$. We assume the billiard to be connected at $x=0$ 
and $x=L$ by two similar point contacts
to leads of width $W$ extended along the $\pm x$-direction.

The beam of electrons injected through a point contact into the billiard has 
cross-section $W$ and transverse momenta in the range $(-p_{W},p_{W})$. The 
number of channels $N\simeq p_{W}W/\hbar$ in the lead is much smaller than 
the number of channels $M\simeq p_{F}L/\hbar$ supported by a typical 
cross-section of the billiard. While $W/L\ll 1$ in general, the ratio 
$p_{W}/p_{F}$ depends on details of the potential near the point contact. 
If $p_{W}/p_{F}\ll 1$ one speaks of a collimated beam. This is typical for 
a smooth potential, while a hard-wall potential typically has 
$p_{W}\simeq p_{F}$ (no collimation). 
%For later use 
%In order to be able to describe the case where the size of the openings in 
%the coordinate and momentum differ strongly 
We define $r_{\rm min}={\rm min}\,(W/L,p_{W}/p_{F})$ and 
$r_{\rm max}={\rm max}\,(W/L,p_{W}/p_{F})$.

The classical phase
space is four-dimensional. By restricting the energy to $E_{F}$ and
taking $x=0$ we obtain the two-dimensional section of phase space shown
in Fig.\ \ref{figure2}. The accessible values of $y$ and $p_{y}$ lie in
a disc-shaped region of area ${\cal A}=Nh$ in this surface of section. 
Up to factors of order unity, the disk has width  $r_{\rm min}$ and
length $r_{\rm max}$ (if coordinate and momentum are measured in units of
$L$ and $p_F$, respectively). In Fig.~2 one has $r_{\rm min}\simeq 
r_{\rm max}$. Each
point in the disc defines a classical trajectory that enters the billiard
(for positive $p_{x}$) and then leaves the billiard either through the
same lead (reflection) or through the other lead (transmission). 
%We use points of different colour (or gray scale) to show transmitted 
%and reflected trajectories.
The points lie
in narrow bands, which we will refer to as ``transmission bands''
and ``reflection bands''.

It is evident from Fig.\ \ref{figure2} that the area $A_{j}$ enclosed
by a typical transmission (or reflection) band $j$ is much less than 
${\cal A}$. For an estimate we consider the time $t(y,p_{y})$ that 
elapses before transmission. 
Let $t_j$ be the dwell time averaged over the starting points $y,p_{y}$ 
in a single band.
The fluctuations of $t$ around the average are of the order of the 
time $t_W\simeq m W/p_W$ to cross the point contact,
which is typically $\ll t_j$. 
As we will see below, the area of the band decreases with $t_j$ as
\begin{equation}
A_{j}\simeq{\cal A}_{0}\exp(-\lambda t_{j})
\;\; 
{\rm if}
\;\; 
t_{j}\gg 1/\lambda , t_W . 
\label{Ajestimate}
\end{equation}
The prefactor ${\cal A}_{0}={\cal A}r_{\rm min}/r_{\rm max}$ 
depends on the degree of collimation. In Ref.\ \cite{Sil02} the 
symmetric case $r_{\rm min}=r_{\rm max}$ was assumed, when 
${\cal A}_{0}={\cal A}$.

\begin{figure}[t]
\includegraphics[width=8cm]
{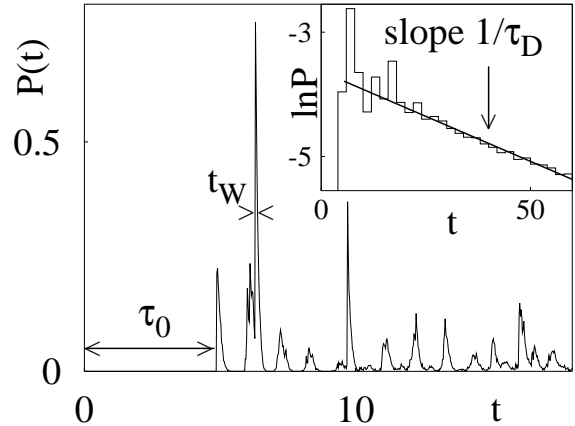}
%{/scratch/marlies/Peter/timegnu.eps}
\caption{
Dwell-time distribution for the billiard of Fig.~1. Electrons at the Fermi
energy are injected through the left lead. Time is in units of $mL/p_F$. 
Inset: the same data on a semilogarithmic scale
with larger bin size of the histogram. Three characteristic time 
scales are seen: $t_W$, $\tau_0$, $\tau_D$.
}
\vspace{-.2cm}
\end{figure}

We now proceed to the construction of fully transmitted scattering
states. To this end we consider a closed contour ${\cal C}$ within a
transmission band $j$. The starting points on the contour define a
family of trajectories that form a flux tube inside the billiard (see
Fig.~1). The semiclassical wave function
\begin{equation}
\psi(x,y)=\sum_{\sigma}\sqrt{\rho_{\sigma}(x,y)}\exp[i
{\cal S}_{\sigma}(x,y)/\hbar] \label{psidef}
\end{equation}
is determined as usual from the action ${\cal S}_{\sigma}$ and density
$\rho_{\sigma}$ that solve the Hamilton-Jacobi and continuity equations
\begin{equation}
|\nabla {\cal S}|^{2}=2m(E_{F}-U),\;\;\nabla\cdot(\rho\nabla 
{\cal S})=0. \label{Hamilton}
\end{equation}
The action is multivalued and the index $\sigma$ labels the different 
sheets. Typically, there are two sheets, one originating from the upper 
half of the contour ${\cal C}$ and one from the lower half.

The requirement that $\psi$ is single-valued as one winds around the
contour imposes a quantization condition on the enclosed area,
\begin{equation}
\oint_{\cal C}p_{y}dy=(n+1/2)h.\label{quantization}
\end{equation}
The increment $1/2$ accounts for the phase shift acquired at the two
turning points on the contour. The quantum number $n=0,1,2,\ldots$
is the channel index. The largest value of $n$ occurs for a contour
enclosing an area $A_{j}$. The number of transmission channels $N_{j}$
within band $j$ is therefore given by $A_{j}/h$, with an accuracy
of order unity. In view of Eq.\ (\ref{Ajestimate}) we have
\begin{subequations}
\label{Njestimate}
\begin{eqnarray}
&&N_{j}\simeq ({\cal A}_{0}/h)\exp(-\lambda t_{j}),\;\;{\rm for}\;\; 
t_{j} < \tau_{E},\label{Njestimatea}\\
&&N_{j}=0,\;\;{\rm for}\;\;t_{j} > \tau_{E}.
\label{Njestimateb}
\end{eqnarray}
\end{subequations}
The time
\begin{equation}
\tau_{E}=\lambda^{-1}\ln({\cal A}_{0}/h)=
\lambda^{-1}\ln(Nr_{\rm min}/r_{\rm max})\label{tauEresult}
\end{equation}
above which there are no fully transmitted channels is the Ehrenfest time
of this problem.

By decomposing one of these $N_{j}$ scattering states into a
given basis of transverse modes in the lead one constructs an 
eigenvector of the
transmission matrix product $tt^{\dagger}$. The corresponding eigenvalue
${\cal T}_{j,n}$ equals unity with exponential accuracy in the semiclassical
limit $n\gg 1$. Because of the degeneracy of this eigenvalue any linear
combination of eigenvectors is again an eigenvector. This manifests itself
in our construction as an arbitrariness in the choice of ${\cal C}$.

We observe in Fig.~1 that the spatial density profile
$\rho(x,y)$ of a fully transmitted scattering state is highly
non-uniform. The flux tube is broad (width of order $W$) at the two
openings, but is squeezed down to very small width inside the billiard. A
similar effect was noted~\cite{Sil02} in the excited states of an Andreev
billiard (a cavity connected
to a superconductor). Following the same argument we estimate
the minimal width of the flux tube as $W_{\rm min}\simeq L\sqrt{N_j/k_F L}$.

The total number
\begin{equation}
N_{0}=\sum_{j}N_{j}=N\int_{0}^{\tau_{E}}P(t)dt\label{N0Pt}
\end{equation}
of fully transmitted and reflected channels is determined by the dwell-time 
distribution $P(t)$ \cite{Bau90}. Fig.~3 shows this distribution in our 
billiard. One sees three different time scales. The narrow peaks 
represent individual transmission~(reflection) bands. 
They consist of an abrupt jump followed by an exponential decay
with time constant $t_W$.
These exponential tails correspond to the borders of the bands, where the 
trajectory bounces many times between the sides of the point contact.
If we smooth $P(t)$
over such short time intervals, an exponential decay with time constant
$\tau_D=\pi\hbar/N\delta$ is obtained (inset). 
The decay starts at the so called ``ergodic
time'' $\tau_0$. There are no trajectories leaving the cavity for $t<\tau_0$.
So the smoothed dwell-time distribution has the form.
\begin{equation}
P(t)=\tau_{ D}^{-1}\exp[(\tau_{0}-t)/\tau_{ D}]\theta(t-\tau_{0}), 
\label{Ptresult}
\end{equation}
with $\theta(t)$ the unit step function.

\begin{figure}
\includegraphics[width=8cm]
{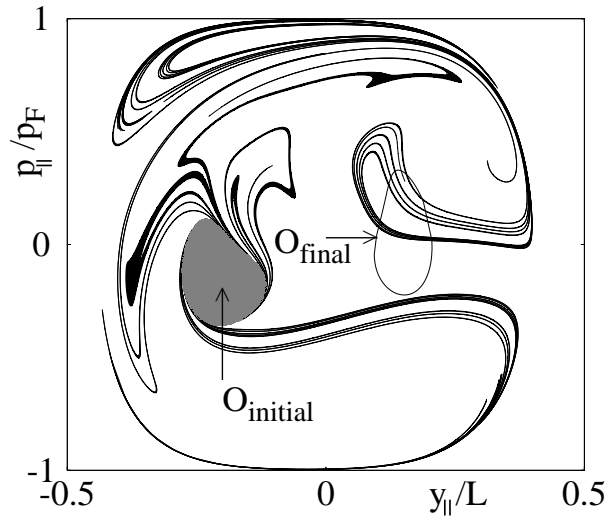}
%{/scratch/marlies/Peter/phasespacemiddlegnulines2.eps}
\caption{
Section of phase space in the middle of the billiard, along line $b$ in 
Fig.~1. 
The subscript $\parallel$ indicates the component of coordinate and momentum
along this line.
Elongated black areas $O_j$ show the positions of the 5-th crossing of the 
injected beam with this surface of section.
The area $O_{\rm initial}$ is the position of the first crossing.
Points inside $O_{\rm final}$ leave the billiard without further crossing 
of line $b$. For times less than the ergodic time $\tau_0$ there is no 
intersection between $O_j$ and $O_{\rm final}$.
}
\vspace{-.2cm}
\end{figure}

In order to find $\tau_0$ we consider Fig.~4, where the section of 
phase space along a cut through the middle of the billiard 
is shown (line $b$ in Fig.~1).
It is convenient to measure the momentum and coordinate along 
$b$ in units of $p_F$ and $L$. 
The injected beam crosses the section for the first time over an
area ${O_{\rm initial}}$
of size $r_{\max}\times r_{\min}=h N/p_F L$.
(Fig.~4 has $r_{\min}\simeq r_{\max}$, but the estimates 
hold for any $r_{\min} < r_{\max} < 1$.)
Further crossings consist of increasingly more elongated areas.
The fifth crossing is shown in Fig.~4.
The flux tube intersects line $b$ in a few disjunct areas ${O}_j$, of 
width $r_{\min}e^{-\lambda t}$ and total length $r_{\max}e^{\lambda t}$. 
(Due to conservation of the integral $\oint {\bf p}\cdot d{\bf r}$ enclosing
the flux tube, the total area $\sum_j {O}_j$
decreases only when particles leave the billiard.)
The typical separation of adjacent areas is $(r_{\max}e^{\lambda t})^{-1}$.
To leave the billiard (through the right contact) without further 
crossing of $b$ a particle should
pass through an area $O_{\rm final}\simeq r_{\max}\times r_{\min}$.
This is highly improbable~\cite{endend} until the separation of the areas ${O}_j$
becomes of order $r_{\rm max}$, leading to 
the ergodic time
\bq\label{ergodic} 
\tau_{0}=\lambda^{-1}\ln r_{\rm max}^{-2}.
\ee
The ergodic time varies from $\tau_0\alt \lambda^{-1}$ for 
$r_{\rm max}\simeq 1$ to $\tau_{0}=\lambda^{-1}\ln(k_F L/N)$ 
for $r_{\rm min}\simeq r_{\rm max}$.
The overlap of the areas  
${O}_j$ and $O_{\rm final}$ is the mapping of the transmission band 
onto the surface of section $b$.
It has an area $p_F L r_{\min}^2e^{-\lambda t}
={\cal A}(r_{\min}/r_{\max})e^{-\lambda t}$,
leading to Eq.~(\ref{Ajestimate}).

Substituting Eq.\ (\ref{Ptresult}) into Eq.\ (\ref{N0Pt}) we arrive at the 
number $N_0$ of fully transmitted and reflected channels,
\begin{eqnarray}
&&N_{0}=
N\theta(\tau_{E}-\tau_{0})
\left[1-e^{(\tau_{0}-\tau_{E})/\tau_{\rm D}}\right],\label{N0result}\\
&&\tau_{E}-\tau_{0}=\lambda^{-1}\ln(N^{2}/k_F L).\label{tauEmintau0}
\end{eqnarray}
There are no fully transmitted or reflected channels if $\tau_{E}<\tau_{0}$, 
hence if $N<\sqrt{k_F L}$. Notice that the dependence of $\tau_{E}$ 
and $\tau_{0}$ separately on the degree of collimation drops out 
of the difference $\tau_{E}-\tau_{0}$. The number of noiseless channels 
is therefore insensitive to details of the confining potential.
An Ehrenfest time $\propto \ln(N^{2}/k_F L)$ has appeared before in 
connection with the Andreev billiard~\cite{Vav02}, but the role of 
collimation (and the associated finite ergodic time) was not considered 
there.

Eqs. (\ref{Njestimate}) and (\ref{Ptresult}) imply that 
the majority of noiseless channels group in
bands having $N_j\gg 1$, which justifies the semiclassical
approximation. 
The total number of these noiseless bands is 
$(N-N_0)/\lambda\tau_D$, which is much less than both $N-N_0$ and $N_0$.
Because of this inequality the relatively short  
trajectories contributing to the noiseless channels are well separated 
in phase space from other, longer trajectories (cf. Fig.~2).

The shot-noise power $S$ is related to the transmission eigenvalues 
by \cite{But90}
\begin{equation}
S=2e\bar{I}g^{-1}\sum_{k=1}^{N}{\cal T}_{k}(1-{\cal T}_{k}),\label{Sdef}
\end{equation}
with $\bar{I}$ the time-averaged current and $g=\sum_{k}{\cal T}_{k}$ 
the dimensionless conductance. The $N_0$ fully transmitted or reflected 
channels have 
${\cal T}_{k}=1$ {or} $0$, hence they do not contribute to the noise. 
The remaining $N-N_{0}$ channels contribute 
at most 1/4 per channel to $Sg/2e\bar{I}$. Using that $g=N/2$ for large 
$N$, we arrive at an upper bound for the noise power $S<e\bar{I}(1-N_{0}/N)$.

For a more quantitative description of the noise power we need to know 
the distribution $P({\cal T})$ of the transmission eigenvalues for the $N-N_0$ 
noisy channels, which can not be described semiclassically. 
We expect the distribution to have the same bimodal form 
$P({\cal T})=\pi^{-1}{\cal T}^{-1/2}(1-{\cal T})^{-1/2}$ 
as in the case $N_{0}=0$~\cite{Jal94}.
This expectation is motivated by the earlier observation that the 
$N_0$ noiseless channels 
are well separated in phase space from the $N-N_0$ noisy ones.
Using this form of $P({\cal T})$ we find that the contribution to 
$Sg/2e\bar{I}$ per noisy 
channel equals $\int_{0}^{1}{\cal T}(1-{\cal T})P({\cal T})d{\cal T}=1/8$, 
half the maximum value. 
The Fano factor $F=S/2e\bar{I}$ is thus estimated as
\begin{subequations}
\label{Fresult}
\begin{eqnarray}
F&=&
{\textstyle\frac{1}{4}},\;\;{\rm for}\;\;N\alt\sqrt{k_{F}L},
\label{Fresulta}\\ 
F&=&
{\textstyle\frac{1}{4}}(k_{F}L/N^{2})^{N\delta/\pi\hbar\lambda}, 
\;\;{\rm for}\;\;N\agt\sqrt{k_{F}L}. \label{Fresultb}
\end{eqnarray}
\end{subequations}

This result should be compared with that of Ref.~\cite{Aga00}:
$F'={\textstyle\frac{1}{4}}(k_{F}L)^{-N\delta/\pi\hbar\lambda}$.
The ratio $F'/F=\exp[(2N\delta/\pi\hbar\lambda)
\ln(N/k_F L)]$ is always close to unity (because ${N\delta/\pi\hbar\lambda}
\simeq N/k_F L\ll 1$). But $F-{\textstyle\frac{1}{4}}$ 
and $F'-{\textstyle\frac{1}{4}}$ are entirely different for
$N\alt\sqrt{k_{F}L}$,
which is the relevant regime in the experiment~\cite{Obe02}.
There the $N$-dependence of the shot noise was fitted as
$F={\textstyle\frac{1}{4}}(1-t_Q/\tau_D)=
{\textstyle\frac{1}{4}}(1-
{\rm constant} \times N)$,
where $t_Q$ is some $N$-independent time.
%, intrinsic for the quantum dot. 
Eq.~(\ref{Fresult}) predicts a more complex $N$-dependence,
a plateau followed by a decrease as $\ln F\propto -N\ln (N^2/k_F L)$,
which could be observable if the experiment extends over a larger range
of $N$.

We mention two other experimentally observable features of the theory 
presented here. The reduction of the Fano factor described by 
Eq.~(\ref{Fresult}) is the cumulative effect of many noiseless bands. 
The appearance of new bands with increasing $N$ introduces a
fine structure in $F(N)$, consisting of a series of cusps
with a square-root singularity near the cusp. The second feature is the
highly nonuniform spatial extension of open channels, evident in Fig.~1,
which could be observed with the STM technique of Ref.~\cite{Top01}.
From a more general perspective the noiseless channels constructed 
in this paper show that the random matrix approach may be used 
in ballistic systems only for sufficiently 
small openings: $N\alt\sqrt{k_FL}$ is required. For larger $N$ the scattering 
becomes deterministic, rather than stochastic, and random matrix theory starts
to break down.

This work was supported by the Dutch Science Foundation NWO/FOM.


\begin{thebibliography}{99}
\bibitem{Bee91} C. W. J. Beenakker and H. van Houten, Phys.\ Rev.\ B {\bf 43},
12066 (1991).
\bibitem{Aga00} O. Agam, I. Aleiner, and A. Larkin, Phys.\ Rev.\ Lett.\ {\bf
85}, 3153 (2000).
\bibitem{Hen02} H.-S. Sim, and H. Schomerus, Phys.\ Rev.\ Lett.\ {\bf
89}, 066801 (2002); R. G. Nazmitdinov, H.-S. Sim, H. Schomerus, and I. Rotter,
Phys. Rev. B {\bf 66}, 241302(R) (2002).
%cond-mat/0208576.
\bibitem{Zas81} G. M. Zaslavsky, Phys.\ Rep.\ {\bf 80}, 157 (1981). 
\bibitem{Obe02} S. Oberholzer, E. V. Sukhorukov, and C. Sch\"{o}nenberger,
Nature {\bf 415}, 765 (2002).
\bibitem{Jal94} R. A. Jalabert, J.-L. Pichard, and C. W. J. Beenakker, 
Europhys.\ Lett.\ {\bf 27}, 255 (1994).
\bibitem{Sil02} P. G. Silvestrov, M. C. Goorden, and C. W. J. Beenakker, 
Phys. Rev. Lett. {\bf 90}, 116801 (2003)
%cond-mat/0208192.
\bibitem{Bau90} 
The dwell-time distribution is defined with a uniform
measure in the surface of section of the lead, so that its integral
is directly proportional to the number of channels.
The mean dwell
time $\tau_{\rm D}=\pi\hbar/N\delta$ was calculated by 
W. Bauer and
G. F. Bertsch, Phys.\ Rev.\ Lett.\ {\bf 65}, 2213 (1990).
\bibitem{endend} Dwell times shorter than $\tau_0$ are improbable but they 
may exist for special positions of the two point contacts.
The probability that a random position permits a
transmitted(or reflected) trajectory of duration $t<\tau_0$
is $\exp[-\lambda (\tau_0-t)]$. 
\bibitem{Vav02} M. G. Vavilov and A. I. Larkin, Phys. Rev. B {\bf 67}, 
115335 (2003).
%cond-mat/0210033.
\bibitem{But90} M. B\"{u}ttiker, Phys.\ Rev.\ Lett.\ {\bf 65}, 2901 (1990).
\bibitem{Top01}  M.~A.~Topinka
%, B.~J.~LeRoy, R.~M.~Westervelt, S.~E.~J.~Shaw, R.~Fleischmann, E.~J.~Heller, 
%K.~D.~Maranowski, and A.~C.~Gossard
{\it et. al.}, Nature {\bf 410}, 183 (2001).

\end{thebibliography}
\end{document}